# Large Anomalous Hall Effect at Room Temperature in a Fermi-Level-Tuned Kagome Antiferromagnet

*Linxuan Song, Feng Zhou, Hang Li, Bei Ding, Xue Li, Xuekui Xi, Yuan Yao, Yong-Chang Lau, Wenhong Wang\**

Dr. Linxuan Song, Feng Zhou, Dr. Hang Li, Dr. Bei Ding, Xue Li, Dr. Xuekui Xi, Dr. Yuan Yao, Dr. Yong-Chang Lau, Prof. Wenhong Wang
Institute of Physics, Chinese Academy of Sciences, Beijing 100190, China
*Corresponding author. Email: wenhong.wang@iphy.ac.cn
Feng Zhou, Dr. Hang Li, Prof. Wenhong Wang
School of Electronics and Information Engineering, Tiangong University, Tianjin 300387, China
Xue Li, Dr. Yong-Chang Lau
University of Chinese Academy of Sciences, Beijing 100049, China

**Abstract**

The recent discoveries of surprisingly large anomalous Hall effect (AHE) in chiral antiferromagnets have triggered extensive research efforts in various fields, ranging from topological condensed-matter physics to antiferromagnetic spintronics, and energy harvesting technology. However, such AHE-hosting antiferromagnetic materials are rare in nature. Herein, we demonstrate that $Mn_{2.4}Ga$, a Fermi-level-tuned kagome antiferromagnet, has a large anomalous Hall conductivity of about 150 $\Omega^{-1}cm^{-1}$ at room temperature that surpasses the usual high values (i.e., 20~50 $\Omega^{-1}cm^{-1}$) observed so far in two outstanding kagome antiferromagnets, $Mn_3Sn$ and $Mn_3Ge$. The spin triangular structure of $Mn_{2.4}Ga$ guarantees a nonzero Berry curvature while generates only a weak net moment of about 0.05 $\mu_B$ per formula unit in the kagome plane. Moreover, the anomalous Hall conductivity exhibits a sign reversal with the rotation of a small magnetic field, which can be ascribed to the field-controlled chirality of the spin triangular structure. Our theoretical calculations indicate that the large AHE in $Mn_{2.4}Ga$ originates from a significantly enhanced Berry curvature associated with the tuning of the Fermi level close to the Weyl points. These properties, together with the ability to manipulate moment orientations using a moderate external magnetic field, make $Mn_{2.4}Ga$ extremely exciting for future antiferromagnetic spintronics.

# 1 Introduction

Kagome-based materials with topologically nontrivial electronic structures have attracted tremendous attention in the past decade in various fields, ranging from topological condensed-matter physics to materials science, and energy harvesting technology.[1-7] Recently, a series of kagome-based ferromagnets have been observed to have anomalous transport response, topological magnetic textures, giant spin-orbit tunability, and correlated topological electronic structures.[8-15] In addition, this trend has also triggered renewed interest in understanding the mechanism of the anomalous Hall effect (AHE) observed in ferromagnetic metals, which is the spontaneous transverse voltage drop induced by a longitudinal electric current.[16] The AHE has long been considered to be proportional to the magnetization of materials, and thus can be seen only in ferromagnets.[17, 18] In contrast, the recent theoretical and experimental advances have demonstrated that the intrinsic AHE is not driven by the magnetization, but by the Berry curvature which is entirely determined by the band topology. It means that, the sizable AHE can be realized not only in ferromagnets, but also in materials without net magnetization, such as spin liquids and antiferromagnets, as long as a nonzero Berry curvature is guaranteed in the momentum-space.[19-22] A key discovery along this direction is the surprisingly large AHE observed at room temperature in the kagome-based hexagonal antiferromagnets $Mn_3Sn$ (space group $P6_3/mmc$), in which all the Mn atoms form a breathing type of kagome lattice in the x-y plane.[23] When cooled below its Néel temperature of 430 K, the Mn magnetic moments form an inverse triangular spin configuration, yielding a nonzero Berry curvature and producing a weak net magnetization with small magnetic anisotropy within the kagome plane. These unique properties make the antiferromagnetic structure of $Mn_3Sn$ easily switchable via various external stimulants, such as magnetic fields, spin-polarized currents and a uniaxial strain.[24-27]

Shortly after the discovery of the large AHE in $Mn_3Sn$, another kagome-based hexagonal antiferromagnet $Mn_3Ge$, isostructural to $Mn_3Sn$, was also reported with a large anomalous zero-field Hall conductivity of approximately 50 $\Omega^{-1}cm^{-1}$ at room temperature.[28, 29] In addition to the large AHE, other unique anomalous thermal, optical and topological transport responses, such as the large anomalous Nernst effect (ANE), the magneto-optical Kerr effect, the planar Hall effect (PHE) and the topological Hall effect, have subsequently been observed in $Mn_3Sn$ and $Mn_3Ge$, arising from the large Berry curvature due to gapped magnetic nodal lines leading to Weyl fermions.[30-36] In particular, a field-induced linear AHE has recently been reported in $Mn_3Sn$, due to spin chirality induced Berry curvature on the Fermi surface by gapping the Weyl nodal line [27]. Nevertheless, the large AHE has thus far only been reported in the above-mentioned two kagome-based noncollinear antiferromagnets $Mn_3Sn$ and $Mn_3Ge$, and more recently in a non-kagome based non-collinear antiferromagnet YbMnBi$_2$.[37] Therefore, the search for new kagome-based noncollinear antiferromagnets with large AHE is of considerable importance, which will broaden the material platform for future antiferromagnetic spintronics as well as for in the search for novel topological phenomena.

Here we report on the first experimental realization of the long-sought-after last member of the Mn$_3$X kagome antiferromagnet family, off-stoichiometric Mn$_3$Ga (Mn$_{2.4}$Ga) in the form of single crystals which shows the large anomalous Hall conductivity by tuning the Fermi level close to the Weyl points. We show that, despite a very small magnetization of ~0.05 μ$_B$ per formula unit in the kagome plane, the single crystal of Mn$_{2.4}$Ga exhibits a strikingly large anomalous Hall conductivity in zero magnetic field of approximately 150 Ω$^{-1}$cm$^{-1}$ at room temperature and approximately 530 Ω$^{-1}$cm$^{-1}$ at 10 K. The angular dependence of the AHE measurements further confirms that the small residual in-plane magnetic moment plays no role in the observed effect, except to control the chirality of the spin triangular structure. Our theoretical calculations demonstrate that the large AHE in Mn$_{2.4}$Ga originates from a significantly enhanced Berry curvature associated with the tuning of the Fermi level close to the Weyl points. Our work demonstrates the great potential of the Fermi-level-tuned Mn$_{3-x}$Ga for future antiferromagnetic spintronics.

## 2 Results and Discussion

The manganese-based binary alloys Mn$_{3-x}$Ga are known to crystallize in different structures such as cubic, tetragonal and hexagonal phases, depending on the annealing temperature and the composition.[38-51] Among them, the hexagonal phase exists in the range of Mn$_{2.35}$Ga-Mn$_{2.8}$Ga and exhibits a layered kagome antiferromagnetic structure with an ordering temperature of $T_N$~430 to 480 K (see Figure S1 in the Supplementary Materials). As shown in **Figure 1a**, the hexagonal-structured Mn$_3$Ga, similar to Mn$_3$Sn and Mn$_3$Ge, consists of two layers of Mn triangles stacked along the c-axis. In each layer, the Mn atoms form a slightly distorted kagome lattice with Ga located at the center of a hexagon. The structure is only stable in the presence of excess Ga randomly occupying the Mn sites.[38] Below its ordering temperature, the Mn moments form a noncolinear triangular antiferromagnetic structure. In this structure, neighboring Mn moments lie in the x-y plane at an angle of 120° (**Figure 1b**), as previously revealed by neutron diffraction studies.[39] It should be noted that, to date, there are only scarce studies on polycrystalline hexagonal Mn$_3$Ga and no report on the properties of single crystals, possibly due to the complicated Mn-Ga binary phase diagram,[52] which hinders the growth of stoichiometric hexagonal Mn$_3$Ga in equilibrium state. To the best of our knowledge, the topological Hall effect and the large piezospintronic effect have only been observed in polycrystalline stoichiometric Mn$_3$Ga in bulk and thin film forms, respectively. [53, 54] In this study, we have successfully grown high-quality single crystals of non-stoichiometric Mn$_{2.4}$Ga (see Figure S2 in the Supplementary Materials). Using single-crystal X-ray diffraction (XRD), it has been confirmed that Mn$_{2.4}$Ga has a hexagonal Ni$_3$Sn-type structure with space group P6$_3$/mmc. The extracted lattice parameters are as follows: $a$ = 5.4135(8) Å, $c$ = 4.3605(9) Å, and $\alpha$=90, $\beta$ = 120°. The temperature dependence of the magnetization M (T) indicates a $T_N$ of 435 K for the single crystals of Mn$_{2.4}$Ga. These parameters are in good agreement with those detected in polycrystalline Mn$_{2.4}$Ga (see Figure S1 in the Supplementary Materials). In addition, high-resolution scanning transmission electron microscopy (STEM) and selected-area electron diffraction (SAED) were performed to further confirm the crystal structure. According to the structural model obtained by single crystal XRD, the zone axes of the SAED patterns in Figure 1b can be identified as [0001], and the crystal orientations are successfully indexed. The high-angle annular dark-field HAADF-STEM image taken along the same zone axes can be perfectly overlaid with the structural model, verifying that the index of the SAED patterns is self-consistent and is that of the identified crystal.

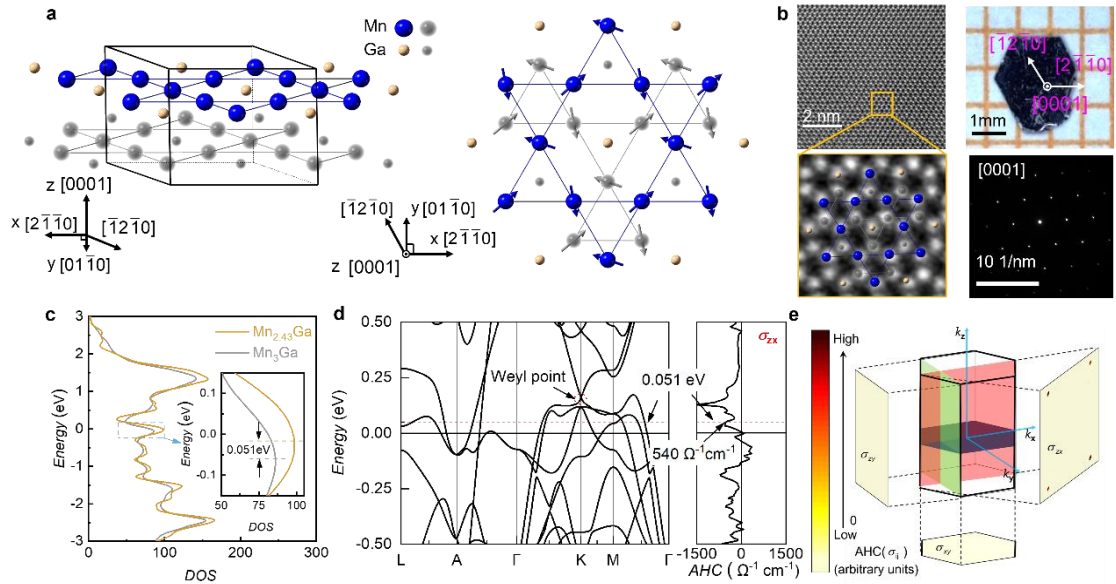

**Figure1. Crystal, magnetic and electronic structures. a.** The crystallographic unit cell of Mn$_3$Ga. In each unit cell, two layers Mn-Ga atoms stuck along z [0001] direction. Small and big spheres represent Ga and Mn atoms respectively. The colored and gray balls represent the top and bottom layer respectively. The kagome lattice in x-y plane is shown in the right panel. The Mn moments of each triangular lattice in the x-y plane form the noncollinear antiferromagnetic configuration. **b.** High-resolution STEM images (left) along the [0001] directions of single crystal Mn$_{2.4}$Ga. The SAED pattern is shown in the bottom right corner to confirm the direction. The optical photograph is shown in top right corner. **c.** Total density of states of the Mn$_3$Ga and Mn$_{2.43}$Ga. **d.** Band structure and anomalous Hall conductivity σ$_{zx}$ of the Mn$_3$Ga. The red dash line shows the Fermi level shift for Ga-rich Mn$_{2.43}$Ga. **e.** First Brillouin zone and momentum-dependent AHC of Mn$_3$Ga. The component of the AHC tensor $\sigma_{ij}$ is shown in the k$_i$-k$_j$ plane, which corresponds to the Berry curvature of all occupied electronic states integrated over the k$_k$ direction (k∥i×j).

## 2.1 Band structures and Berry curvatures

We have performed the electronic band structure calculations of Mn$_3$Ga, which are consistent with previous calculations.[55] As shown in **Figure 1c**, the calculated total density of states of the Mn$_3$Ga and Mn$_{2.43}$Ga shows a peak shift of about 0.051 eV, which indicates the Fermi level ($E_F$) will move up. The band structures show little difference expect for the Fermi level shifts between the Mn$_3$Ga and Mn$_{2.43}$Ga (see Figure S4 in the Supplementary Materials). In both cases, when spin-orbit coupling (SOC) is included, the nodal line is gapped, resulting in two pairs of Weyl points and a nonzero Berry curvature. The strength of the Berry curvature depends on the position of the $E_F$, and reaches its maximum when the $E_F$ is near the Weyl points. From these band structure calculations, we highlight that, although the stoichiometric Mn$_3$Ga has a very small AHC, a large AHC may be produced in the non-stoichiometric Mn$_{2.4}$Ga due to the $E_F$ shift being close to the Weyl points. For comparison, in **Figure 1d**, we show the calculated AHC of Mn$_3$Ga. One can see that the AHC for z-x ($\sigma_{zx}$) component is nearly zero for Mn$_3$Ga, whereas this value reaches up to 540 $\Omega^{-1}$cm$^{-1}$ (for Mn$_{2.43}$Ga, due to the $E_F$ shifted up approaching to the Weyl points). Indeed, this value of 540 $\Omega^{-1}$cm$^{-1}$ in Mn$_{2.43}$Ga is very large and exceeds the other two Kagome-based antiferromagnets Mn$_3$Sn and Mn$_3$Ge. Here, $\sigma_{ij}$ corresponds to the AHC when current flows along the *j* direction, and thereby results in a Hall voltage along the *i* direction.

The anisotropy of AHC for Mn$_3$Ga, i.e., $\sigma_{zx}$ in the x-z plane, $\sigma_{zy}$ in the y-z plane and $\sigma_{xy}$ in the x-y plane are further calculated, and the results are shown in Figure 1e. We found that only the $\sigma_{zx}$ shows the stronger AHC, while the $\sigma_{zy}$ and $\sigma_{xy}$ is nearly zero, which is consistent with the previous calculations of the Mn$_3$Sn and Mn$_3$Ge demonstrated that $\sigma_{zy}$ and $\sigma_{xy}$ should be zero due to the mirror symmetry of antiferromagnetic structure with respect to the x-z plane.[29, 55] It should be note that, in Mn$_3$Ge and Mn$_3$Sn, experimental non-zero value of $\sigma_{zy}$ can be attributed to the perturbation of the mirror symmetry caused by the non-zero moment in the kagome plane. In this sense, we predicted that, similar to Mn$_3$Ge and Mn$_3$Sn, the non-stoichiometric Mn$_{2.4}$Ga may show a large AHE and also a strong anisotropy AHC experimentally.

## 2.2 Large anomalous Hall effect

We first provide clear evidence of the large AHE observed in single crystals of Mn$_{2.4}$Ga. All the transport measurements were performed on Hall devices with about 15 μm × 7 μm ×1μm channels for longitudinal and Hall resistivity measurements (see Experiment Section and Figure

S5 in Supplementary Materials). **Figure 2a** shows the magnetic field $H$ dependence of the Hall resistivity $\rho_{zx}$ for the current $I$ along [2-1-10] and $H$ parallel to [01-10], namely, H is parallel to the kagome plane (which we label as configuration I). Obviously, the $\rho_{zx}$ curve exhibits a clear hysteresis loop with a sharp jump and saturates in modest magnetic fields, attaining a large saturation value of ~2 μΩ·cm at 300 K and ~5.9 μΩ·cm at 10 K. In other words, a large remanent Hall effect at zero field is observed in configuration I and suggests that these Hall devices are mostly a single AFM domain. To make comparison with theory later, here we take the x, y and z coordinates along [2-1-10], [01-10] and [0001], and estimate the anomalous Hall conductivity (AHC). As shown in **Figure. 2d**, the AHC for the configuration I ($\sigma_{zx}$), calculated from $\sigma_H = -\rho_H/\rho^2$, (where $\rho$ is the longitudinal resistivity), exhibits a large value of ~150 $\Omega^{-1}$ cm$^{-1}$ at 300 K, and it becomes particularly enhanced at low temperatures and reaches ~530 $\Omega^{-1}$ cm$^{-1}$ at 10 K, which is quite high, especially for an antiferromagnet. Remarkably, the large AHC of ~150 $\Omega^{-1}$cm$^{-1}$ at 300 K, to the best of our knowledge, surpasses the high values observed in Mn$_3$Sn ~20 $\Omega^{-1}$cm$^{-1}$ and Mn$_3$Ge ~50 $\Omega^{-1}$cm$^{-1}$.

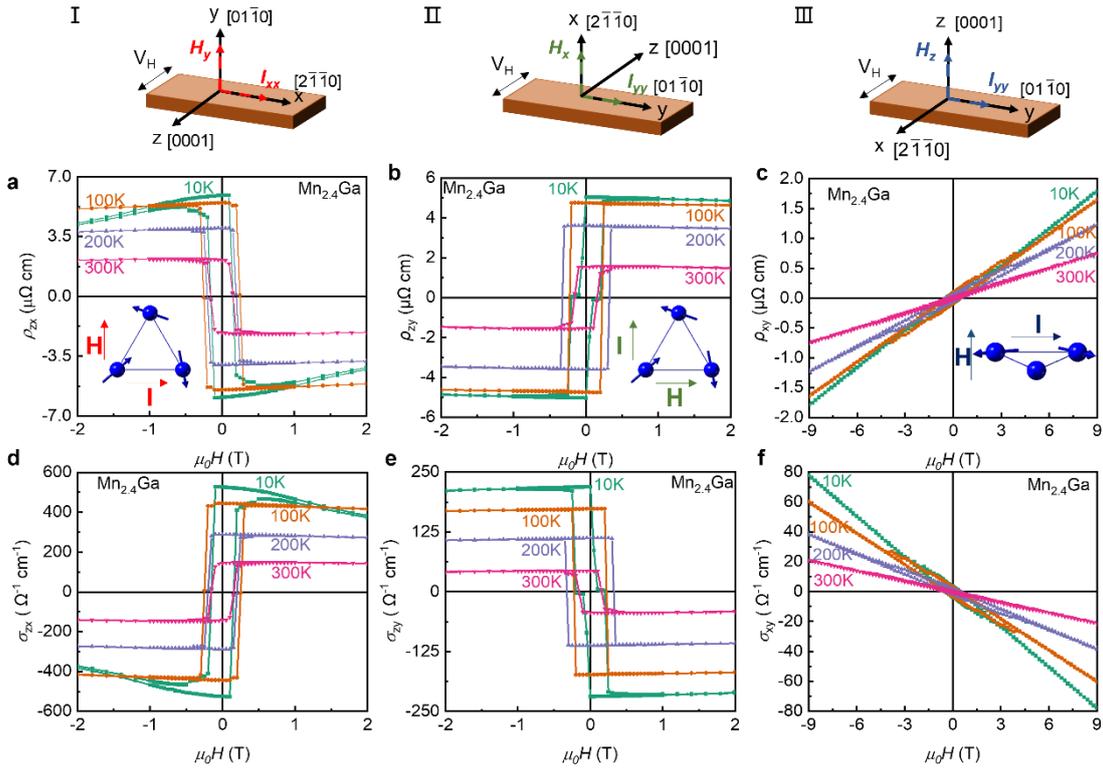

**Figure 2. Magnetic field dependence of anomalous Hall effect AHE in Mn$_{2.4}$Ga. a, b, c,** Field dependence of the Hall resistivity $\rho_H$ and **d, e, f**, AHC measured with three different current and magnetic field configurations: the magnetic field $H // $ y with the electric current $I // $ x (configuration I), $H // $ x, $I // $ y (configuration II) and $H // $ z, $I // $ y (configuration III), at various temperatures. The configurations are illustrated above the corresponding data. The triangular magnetic structures were shown in the inserts to further indicate the measurement configurations.

To study whether the experimental AHE has an anisotropy, as predicted by theory, we have measured $\rho_{zy}$ with $I$ along [01-10] and with $H$ parallel to [2-1-10]. In this configuration, H is still parallel to the kagome plane. As shown in **Figure 2b**, the $\rho_{zy}$ is slightly smaller than configuration I (~5.1 μΩ·cm at 10 K and 1.6 μΩ·cm at 300 K). In contrast, $\sigma_{zy}$ is significantly reduced down to about 220 $\Omega^{-1}cm^{-1}$ at 10 K in configuration II compared to configuration I, while a modest value of $\sigma_H$ = 45 $\Omega^{-1}cm^{-1}$ at 300 K is found for configuration II (**Figure 2e**). This is again very large for an AFM and comparable to those values found so far in antiferromagnets. For comparison, in configuration III, we apply $I$ along [2-1-10] and $H$ parallel to [0001], namely $H$ is perpendicular to the kagome plane. In this configuration, both $\rho_{xy}$ and $\sigma_{xy}$ show no hysteresis but only a nearly linear field dependence at all temperatures, as shown in **Figure 2c** and **2f**, respectively. The more detailed results can be found in Figure S6-S9 in the Supplementary materials.

Notably, the experimental results are highly consistent with the theoretical predictions: large AHE are observed in $\sigma_{zx}$ and $\sigma_{zy}$, but not in $\sigma_{xy}$. Therefore, the nonzero Berry curvature related to the noncollinear antiferromagnetism is believed to be responsible for AHE. It should be noted that an artificial AHE signals could be induced by inhomogeneous Hall signals due to thickness variation, defects, and interface modification. However, these extrinsic factors can be safely excluded in this work. More importantly, the reliability of the experimental protocols was further confirmed by the Mn$_3$Sn single-crystals devices fabricated using the same microfabrication method. A detailed validity analysis of the AHE can be found in Figure S10 and S11 in the Supplementary Materials. Furthermore, we experimentally verified that the sign of the AHC is the same for non-stoichiometric Mn$_{2.4}$Ga and Mn$_3$Sn, which is in line with theoretical calculations [30, 55].

**2.3 In-plane weak ferromagnetism**

To confirm that the observation of a large AHE originates from the noncolinear antiferromagnetic spin structure in the kagome plane, we have performed magnetization measurements with the field parallel to different crystallographic directions. **Figure 3a** shows temperature dependence of magnetization of Mn$_{2.4}$Ga with the field parallel to x, y, and z directions, respectively. The ordering temperature of $T_N$~435 K is consistent with data for polycrystals. In addition, the magnetization curves show anisotropic hysteresis similar to that found for the Hall effect. For example, as shown in **Figure 3b**, the M-H loops measured with the field parallel to the in-plane [01-10] (y) direction at temperatures between 10K and 400K shows clear hysteresis, indicating that a weak spontaneous magnetization of 0.002-0.05 μ$_B$ per formula unit. When the field is applied parallel to the in-plane [2-1-10] (x) direction, as shown in **Figure 3c**, the magnetization is almost isotropic in the Kagome plane and the zero field moment is slightly smaller (0.04 per formula unit at 10 K). A clear magnetic hysteresis is found in both cases, with coercive fields ($H_C$) of less than 0.05T). Although the Mn moments are expected to lie only in the ab plane, when the field is applied parallel to the out-of-plane [0001] (z) direction, we observe a zero field moment that is an order of magnitude less than 0.007 $\mu_B$ per formula unit at 10 K. This indicates that there could be a very small tilting of the Mn spins toward the c-axis. The M-H loops measured at 300 K for different field orientations are similar to the 10 K data, and magnetization shows almost linear dependence behavior at higher fields (**Figure 3d**).

Based on our magnetization data, it is evident that the present kagome-based $Mn_{2.4}Ga$ antiferromagnet has a weak in-plane ferromagnetism and nearly zero net magnetic moment along [0001] (z) direction. We thus argue that the weak in-plane ferromagnetism occurs due to the geometrical frustration of the Mn moments, which are arranged in a triangular noncollinear configuration in the x-y plane. Such configuration is robust against an external field of up to 16T (inset Figure 3d). Hence, it can be concluded that the large AHE observed in the field parallel to the kagome plane (Figure. 2a and 2b) must be related to the triangular noncollinear spin structures. In this manner, the magnetization results are in line with the AHE measurements. The observation of a large AHC in the xz and yz planes, while a very small effect in the x-y plane is also consistent with symmetry analysis.

It should be pointed out that all antiferromagnet kagome $Mn_3X$ alloys are off-stoichiometry triangular antiferromagnet. The hexagonal $Mn_3Ga$ can be only stabilized with the Ga-rich compositions,[39] while the $Mn_3Sn$ and $Mn_3Ge$ can be only obtained with the excessive Mn.[30, 56] The neutron diffraction experiments have confirmed the noncollinear antiferromagnetic structure for these alloys, assuming that the excess atoms randomly occupy the deficiencies.[39, 57] In $Mn_{2.4}Ga$, about 5.9% Mn vacancies were occupied by Ga atoms. While, in $Mn_3Sn$, about 6%~9% Sn atom sites are occupied by Mn atoms,[30, 36] and in $Mn_3Ge$ about 3.6%~13% Ge atom sites are occupied by Mn atoms [28, 36]. However, the polarized neutron diffraction experiments indicate that the weak magnetism in $Mn_3Sn$ mainly originates from the triangular structure and not the excess concentration of Mn which can induce the uncompacted magnetic state.[58] This finding is also supported by the M-H curves of the $Mn_3X$ alloys, which demonstrate that the weak ferromagnetism solely appears in the kagome plane. Most importantly, experiments and theoretical calculations have confirmed that weak ferromagnetism does not play a role in the AHE in $Mn_3Sn$ [55] and $Mn_3Ge$.[29] Though the theoretical calculations have already confirmed this point in $Mn_{2.4}Ga$, we also performed the following experiments to verify it.

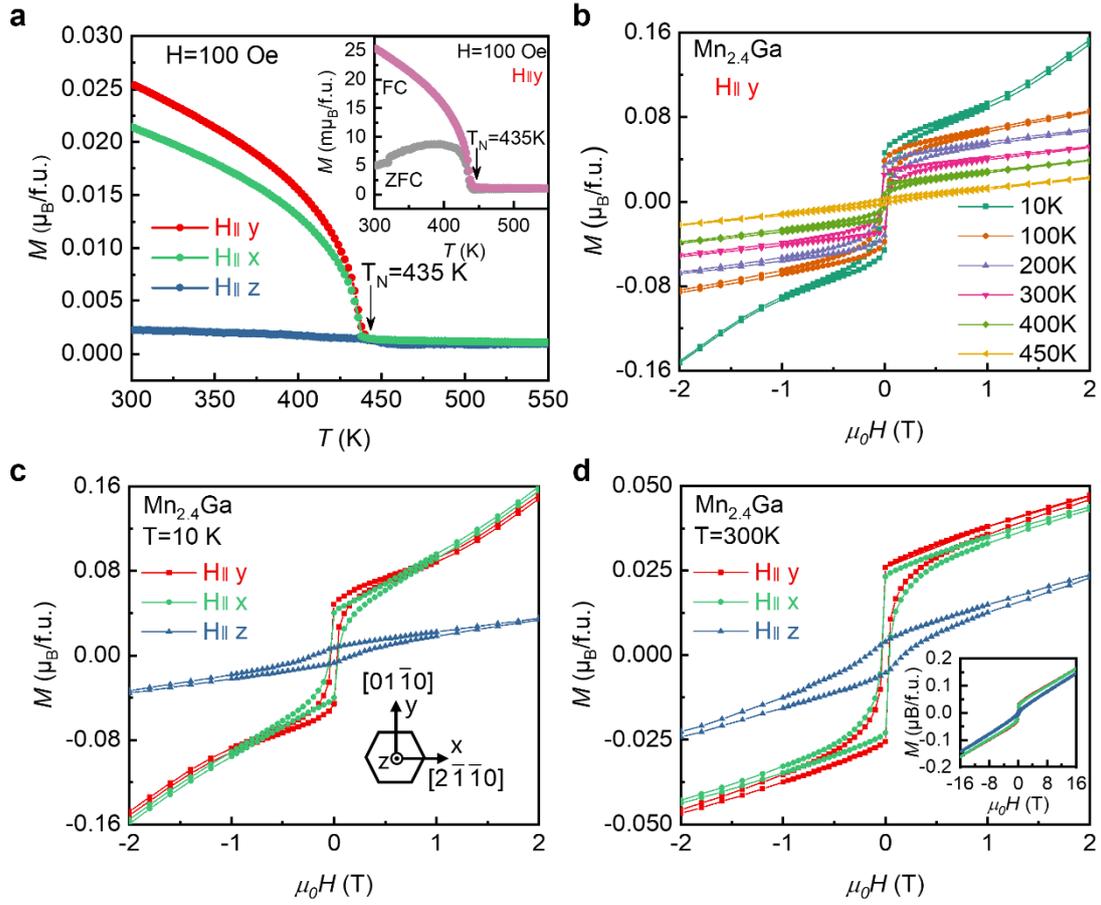

**Figure 3. Magnetic properties of Mn$_{2.4}$Ga. a.** The temperature dependence of the magnetization (M-T) with 100 Oe applied along y, x, z axes respectively. The inset shows the zero field cooling and the field cooling M-T curves with 100 Oe applied along y axis. **b.** Magnetic field dependence of the magnetization (M-H) with field $H // $ y at various temperatures from 10 K to 400 K. **c, d,** Field dependence of the magnetization with magnetic field applied along y, x, z, at 10 K and 300 K respectively. Inset in **d** shows the high magnetic field M-H curves up to 16 T with magnetic field applied along y, x, z, at 300 K.

## 2.4 Angular dependence AHE

To confirm that the weak in-plane ferromagnetism observed in Mn$_{2.4}$Ga does no effect on the large AHE, we performed angular dependence of AHE measurements for various configurations of magnetic field *H* and current *I*. Below, we will demonstrate how drastically the AHC in Mn$_{2.4}$Ga depends on the measurement configuration. **Figure 4a** shows the configuration I, in which the current was applied along x-axis and the Hall voltage $V_H$ is measured within the z-x plane. In this configuration, the initial direction of the magnetic field for $\theta = 0$ is along y. The field was then rotated within the y-z plane. Interestingly, the value of $\sigma_{zy}$ remains constant until $\theta = 90°$, where it abruptly changes sign. With increasing $\theta$, a similar sign change occurs again at $\theta = 270°$. In configuration II shown in **Figure 4b**, we applied the current along y with the Hall voltage $V_H$ measured in the z-y plane, and the field is rotated within the x-z plane. In configuration I and configuration II, the rotating field is always perpendicular to the current and the x-y Kagome plane. We observed that the angular dependence of the $\sigma_{zy}$ follows the same trend as that of configuration I. In both cases, a nearly constant AHC was obtained as long as a small field lies parallel to the in-plane triangular spin structure, with the exception of the abrupt sign change at $\theta = 90°$ and 270°. This behavior can be explained as follows: with increasing $\theta$, although the in-plane magnetic field gradually decreases, a small component of the field remains parallel to the kagome planes up to $\theta = 90°$. This small in-plane field is capable of preserving the chirality of the triangular spin structure in a fixed direction. However, at $\theta > 90°$, the component of in-plane field changes its direction, resulting in a change in the chirality of the spin structure, which causes the sign change in the AHE. A similar phenomenon occurs at $\theta = 270°$ in both configuration I and configuration II. In contrast, in configuration III shown in **Figure 4c**, the field was rotated within the z-x plane, and the Hall voltage $V_H$ was measured in the x-y plane with $I$ along y. In this configuration, a component of the magnetic field always remains parallel to the kagome plane, except for $\theta = 0$ and 180°. However, the AHC measured in the x-y plane ($\sigma_{xy}$) is always negligibly small regardless of the field direction. Notably, a small nonvanishing value of $\sigma_{xy}$ may appear from a slight canting of the spins along the z (c) direction.

Furthermore, such small in-plane field-induced change of chirality can persist up to room temperature. In Figure S12, we show the angular dependence of $\sigma_{zy}$ measured at various temperatures with a fixed field of 1 T in case of configuration I. With increasing of temperature, the $\sigma_{zy}$ gradually decreases and can be attributed to the temperature disturbance on the spin structure. Notably, the angular dependence AHC have also been observed in $Mn_3Ge$ [29]. From these angle-dependent measurements, we can conclude that the small residual in-plane ferromagnetic component, as well as the applied magnetic field, plays no role in the observed large AHE in $Mn_{2.4}Ga$.

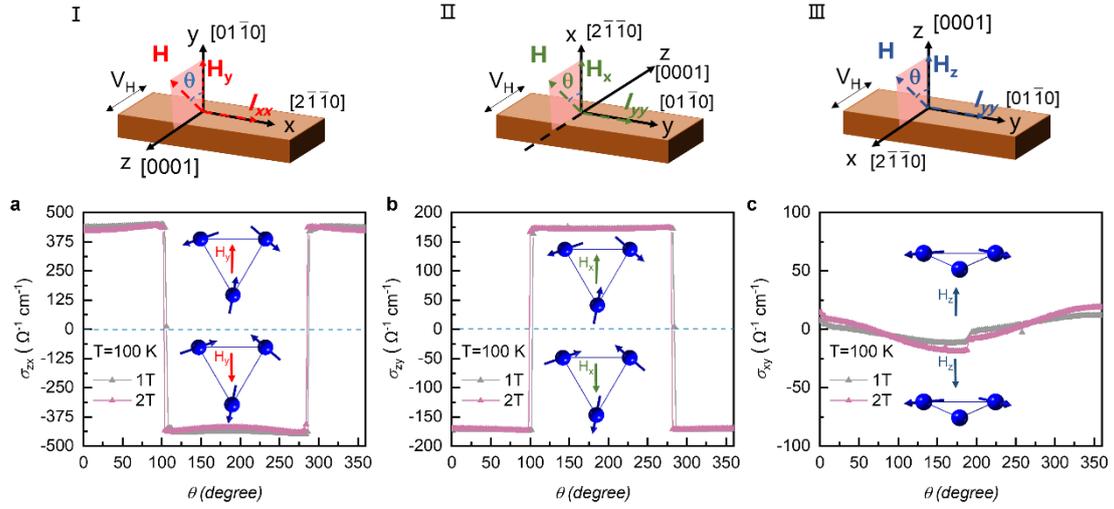

**Figure 4. Angular dependence of AHE for single crystal Mn$_{2.4}$Ga.** Angular ($\theta$) dependence of the AHC in three field and current configurations which are shown by the schematic diagram above the corresponding data. The field rotation was performed in y-z plane for *I* along x (configuration I) **a**, x-z plane for *I* along y, (configuration II) **b**, x-y for *I* along y (configuration III) **c**, respectively. The 1 T and 2 T magnetic fields were applied at 100 K. $\theta$=0° corresponds to the magnetic field oriented along y, x, z respectively. The insets show the changes of triangular magnetic structure chirality induced by the direction of magnetic field component along the y, x, z axes.

## 2.5 Comparison of Mn$_{2.4}$Ga with other AHE materials

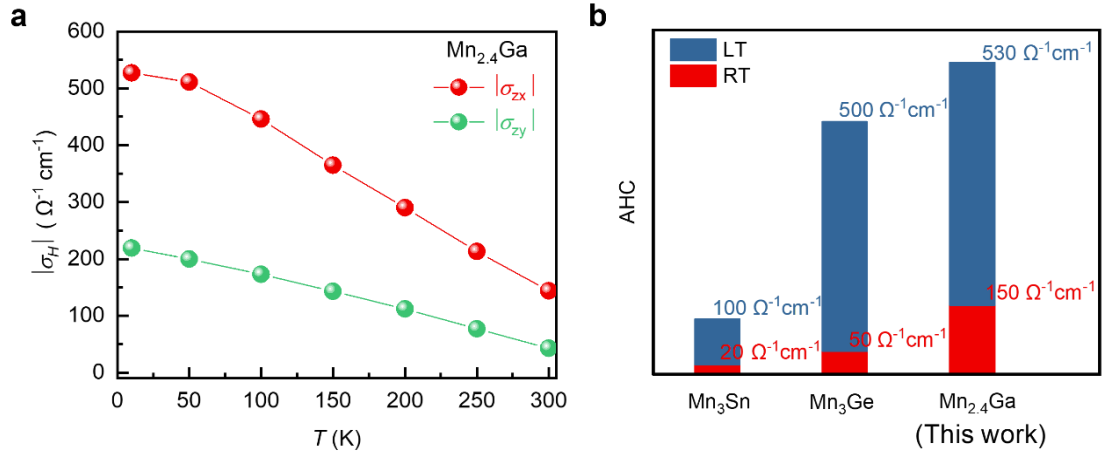

**Figure 5. Summary of anomalous Hall conductivity of the Mn$_{2.4}$Ga. a.** Temperature dependence of the AHC with two field and current configurations: *H //* y, *I //* x; H // y, *I //* x, respectively. **b.** The maximum AHC of the Mn$_3$X alloy (Sn, Ge, Ga) [23, 29] at room temperature (RT) and low temperature (LT), respectively.

The zero-field AHC at various temperatures was extracted from the magnetic field dependance of the AHC by extrapolating the high magnetic field part back to zero field. As shown in **Figure 5a**, with decreasing temperature, the AHC shows the weaker dependence on the temperature, comparing to the Mn$_3$Sn and Mn$_3$Ge. Interestingly, the AHE caused by the Berry curvature should not depend on temperature, such as Fe$_3$Sn$_2$ [59] and Co$_3$Sn$_2$S$_2$. [11] However, for the Kagome antiferromagnets the AHE typically shows strong temperature dependent, including this work (Mn$_3$Ga) and previous work on Mn$_3$Sn [23] and Mn$_3$Ge. [29] Although it is clear now that the Berry curvature generated by the noncollinear AFM structure is crucial for the observed AHE in these compounds, to the best of our knowledge, the origin of its temperature dependence remains an open question. However, the temperature dependent intrinsic AHE has been discussed in the Ni films, which is related to the magnetocrystalline anisotropy changing with the temperature which induced the Fermi energy passing through small band gaps caused by spin-orbital coupling[60]. We speculate that due to the temperature perturbation noncollinear antiferromagnetic structure, the intrinsic Berry-curvature shows a temperature dependent behavior, which results in the temperature dependent intrinsic AHC. The AHC reaches a value of $\sigma_{zx}$ ~530 $\Omega^{-1}$ cm$^{-1}$ and $\sigma_{zy}$ ~220 $\Omega^{-1}$ cm$^{-1}$ at 10K in x-z and y-z plane, respectively, which is quite high, especially for an antiferromagnet (see Table S2). More importantly, the large AHC of $\sigma_{zx}$ ~150 $\Omega^{-1}$ cm$^{-1}$ at 300K are remarkable because it surpasses the highest values observed in Mn$_3$Sn (~20 $\Omega^{-1}$cm$^{-1}$) and Mn$_3$Ge (~50 $\Omega^{-1}$cm$^{-1}$) at room temperature (**Figure 5b**).

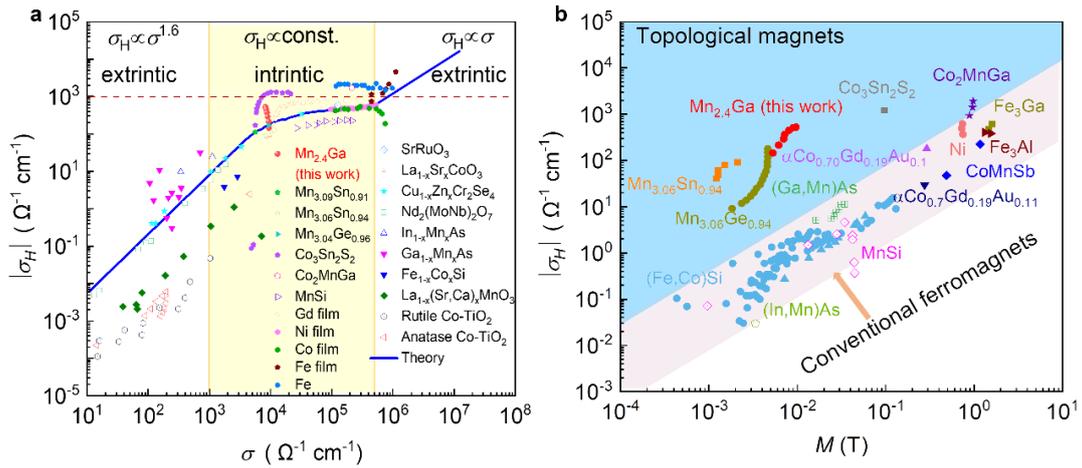

Figure 6. **Comparison of Mn$_{2.4}$Ga with other AHE materials. a.** Universal scaling relation between the Hall conductivity $\sigma_H$ and the longitudinal conductivity $\sigma$ of the Mn$_{2.4}$Ga (10K to 300K) plotted together with various ferromagnets and antiferromagnets including transition metals (Ni, Gd, Fe and Co thin films), [61] Fe single crystals, [61] Co$_3$Sn$_2$S$_2$, [11] Mn$_{3.06}$Sn$_{0.94}$, [23] Mn$_{3.09}$Sn$_{0.91}$, [30] Mn$_{3.04}$Ge$_{0.96}$, [36] MnSi, [62] Fe$_{1-x}$Co$_x$Si, [62] perovskite oxides SrRuO$_3$, La$_{1-x}$Sr$_x$CoO3, La$_{1-x}$(SrCa)$_x$MnO$_3$, [63] chalcogenide spinels Cu$_{1-x}$Zn$_x$Cr$_2$Se$_4$, [63] pyrochlore Nd$_2$(MoNb)$_2$O$_7$ [63] and magnetic semiconductor Ga$_{1-x}$Mn$_x$As, In$_{1-x}$Mn$_x$As, anatase-Co-TiO$_2$, rutile-Co-TiO$_2$. [63] The yellow shaded region marks the moderately dirty regime ($3\times10^3 \leq \sigma \leq 5\times10^5$ $\Omega^{-1}$cm$^{-1}$), in which the $\sigma_H$ is predominated by the intrinsic Berry phase contribution and nearly constant with $\sigma$. The super clean region of $\sigma \geq 5\times10^5$ $\Omega^{-1}$cm$^{-1}$, $\sigma_H$ is mainly dominated by the extrinsic skew scattering. The poorly conducting regime $\sigma \leq 3\times10^3$ $\Omega^{-1}$cm$^{-1}$, the $\sigma_H$ also originates from the extrinsic contribution, shows $\sigma_H \propto \sigma^{1.6}$ relation. The solid line presents the theoretical prediction. [64] **b.** Full logarithmic plot of the magnetization (M) dependence of the AHC of Mn$_{2.4}$Ga, reported topological magnets [23, 30, 65] and conventional ferromagnets. [18, 36, 66] The red shades mark the regions in which $\sigma_H$ is linearly related to M for conventional ferromagnets. The topological magnets that surpass this linear relation are located in the blue shaded region.

To further scale the transport properties of the Mn$_{2.4}$Ga, we have summarized the relation between the Hall conductivity $\sigma_H$ and the longitudinal conductivity $\sigma$ of the AHE materials (**Figure 6a**). The longitudinal $\sigma$ is in the range of $5 \times 10^3$ $\Omega^{-1}$ cm$^{-1}$ indicating that Mn$_{2.4}$Ga is in the "moderately dirty" regime where the AHE is mainly governed by the intrinsic mechanism like Mn$_3$Sn and Mn$_3$Ge. This is also in line with the theoretical calculations. [36, 55] For topological antiferromagnets, in addition to the large AHE, other topological transport properties like ANE, positive longitudinal magnetoconductivity and PHE have all been observed in Mn$_3$Sn and Mn$_3$Ge. These properties are attributed to the large Berry curvature stemming from the Weyl points, which is with not related to the net magnetization. [36] It can be seen that the Mn$_{2.4}$Ga shows an obviously dependent on the $\sigma$. The $\sigma$ is temperature dependent. As we mentioned above, the temperature perturbation noncollinear antiferromagnetic structure will result in the temperature dependent intrinsic AHC as well as the $\sigma$ dependent intrinsic AHC. On the other hand, the magnetization dependence of the spontaneous Hall conductivity $\sigma_H$ of the Mn$_3$X alloys and other ferromagnetic materials are plotted in **Figure 6b**. We note that the Mn$_{2.4}$Ga presented in this work is located in the blue region where the $\sigma_H$ is beyond the linear relation for the conventional ferromagnets (pink region). Besides, the well-known topological materials Co$_3$Sn$_2$S$_2$ and Co$_2$MnGa are also in this regime which have been revealed the evidence of the Weyl fermions by ARPES, [11, 67, 68] as well as the Mn$_3$Sn. [8] Moreover, the large positive magnetoconductivity of Mn$_3$Ge indicate that the Mn$_3$Ge is also a topological material. [36] The present results also suggest that kagome-based antiferromagnets, including but not limited to the Mn$_3$X (X=Sn, Ge, Ga) family, are worthy of further investigation for large AHE.

## 3. Conclusion

We have demonstrated on the first experimental realization of the long-sought-after last member of the Mn$_3$X kagome antiferromagnet family, the off-stoichiometry Mn$_3$Ga in the form of single crystals, which shows large AHE at room temperature. By tuning the Fermi level close to the Weyl points, we found the non-stoichiometric Mn$_{2.4}$Ga single crystal exhibits strikingly large anomalous Hall conductivity in zero field of approximately 150 $\Omega^{-1}$cm$^{-1}$ at room temperature and approximately 530 $\Omega^{-1}$cm$^{-1}$ at 10 K, despite a very small magnetization ~0.05$\mu_B$ per formula unit in the kagome plane. The angular dependence of the AHE measurements further confirms that the small residual in-plane magnetic moment has no role in the observed effect except to control the chirality of the spin triangular structure. Our theoretical calculations demonstrate that the large AHE in Mn$_{2.4}$Ga originates from a significantly enhanced Berry curvature associated by tuning the Fermi level close to the Weyl points. In addition, we have fabricated the polycrystals of hexagonal structures over the range of $Mn_{2.35}Ga$-$Mn_{2.8}Ga$ and exhibits a layered kagome antiferromagnetic structure with a tunable magnetic and AHE (see Figure S13 in Supplementary Materials), which demonstrates the efficient Fermi level engineering via off-stoichiometric substitutional alloying. We thus suggest that, by changing the composition of Mn, a precise tuning of the $E_F$ may be realized, which will further enhance the value of AHC at room temperature. In this sense, the Fermi-level-tuned $Mn_{3-x}Ga$ can be a new paradigm of the field-controlled antiferromagnetic family with promising prospects for future antiferromagnetic spintronics.

## 4. Experimental Section

*Sample preparation:* Single crystals of $Mn_{2.4}Ga$ were prepared by Pb flux method. Firstly, the bulk polycrystalline $Mn_{2.4}Ga$ were prepared by arc melting with high-purity Mn (99.99 wt. %) and Ga (99.99 wt. %) under an argon atmosphere. Secondly, polycrystalline sample and Pb with the molar ratio of Mn:Ga:Pb=2.4:1:22.5 were placed in a $Al_2O_3$ crucible which was sealed in quartz tubes later. Then the quartz tubes were put into an electric furnace, were slowly heated to 1423 K and held on for 2 days. Then the quartz tube was cooled down to 1023 K at a rate of 4 K/h to reach the low temperature phase region. After that the quartz tubes were cooled down to the 953 K at a slower rate of 2 K/h for the growth of single crystals. Finally, to obtain the isolated single crystals, the quartz tubes were moved quickly into a centrifuge to separate the Pb flux at 953 K.

The $Mn_{3-x}Ga$ polycrystals were prepared by the melt-spinning method. Details can be found in our previous work [50].

The single crystals of $Mn_3Sn$ were prepared by Sn self-flux method. Pieces of Mn (99.99%) and Sn (99.99%) were mixed with an atomic ratio of Mn:Sn = 7:3 and placed in an alumina crucible. The crucible was sealed in a quartz tube under vacuum. The quartz tubes were slowly heated to 1423 K and held on for 1 days. Then the quartz tubes were cooled down to 1273 K at a rate of 25K/h. After that the quartz tubes were cooled down to the 1173 K at a rate of 1 K /h. Finally, to obtain the isolated single crystals, the quartz tubes were moved quickly into a centrifuge to separate the Sn flux at 1173 K.

*Structural and composition characterizations:* The crystal structure and orientation of the as-grown single crystals were determined using single-crystal X-ray diffraction (RAPID, Rigaku) at room temperature. The lattice parameters were obtained by Rietveld refinement. All the samples were shown to be in single phase, with lattice parameters close to those found in previous polycrystals work. According to the energy dispersive X-ray analysis with a scanning electron microscope, the composition of single crystals was determined to be $Mn_{2.4}Ga$.

*Magnetization measurements:* The magnetization measurements on orientated samples were conducted using a commercial SQUID magnetometer (MPMS, Quantum Design) equipped with the high temperature module in the temperature range of 2K–800 K. The high magnetic field measurements were conducted using a Physical Property Measurement System (PPMS, Quantum Design) with a maximum applied magnetic field 16 T.

*Device fabrication:* First**,** thin flakes with different orientations were fabricated from the single crystal of $Mn_{2.4}Ga$ using a focused ion-beam system (Helions PFIB CXe). Second, the thin flakes were then transferred to a copper chip by the tungsten needle to further reduce the thickness of the flakes to ~1μm. Third, the thin flakes were transferred to the silicon wafer with pre-made electrodes. Finally, the electrodes and the flake were connected by the plating Pt. $Mn_3Sn$ devices were fabricated using similar procedure.

*Transport measurements:* The longitudinal resistance and Hall voltage signals of devices were measured simultaneously in standard five-probe geometry. Measurements at temperatures of 10–300 K were performed in the PPMS system (Quantum Design). In order to eliminate the influence of probes misalignment, the Hall contributions to the longitudinal resistivity and vice versa were eliminated by adding and subtracting the resistivity data taken at positive and negative magnetic fields.

*First principles calculations:* Calculations were conducted using density functional theory (DFT) implemented in the Vienna ab-initio simulation package [69] (VASP) code. The exchange correlation functional is the Generalized-Gradient-Approximation [70] (GGA) of the Perdew-Burke-Ernzerhof [71] (PBE) functional. The cutoff energy was set as 450 eV, energy and force convergence criteria set as $10^{-5}$ eV and 0.01 eV Å$^{-1}$. Spin-orbit coupling (SOC) was included in all calculations. $Mn_{2.43}Ga$ was built by considering 3×3×1 $Mn_3Ga$ supercell and replacing three Mn atoms with Ga randomly. The Mn magnetic moment is about 3 $\mu_B$ for $Mn_3Ga$ and 2.4 $\mu_B$ for $Mn_{2.43}Ga$, these calculated values agree well with the experimental resutls. The anomalous Hall conductivity and Berry curvature were calculated in Wannier90 [72, 73] and Wanniertools package [74].

**Supporting Information**

Supporting Information is available from the Wiley Online Library or from the author.

**Conflict of Ineterest**

The authors declare that they do not have any conflicts to disclose.

**Acknowledgements**

This work was supported by the National Key R&D program of China (No. 2022YFA1402600), The Beijing Natural Science Foundation (Grant No. Z230006) and National Natural Science Foundation of China (Grants No. 12204347, No. 12274438, No. 12274321, and No. 12074415). A portion of this work was carried out at the Synergetic Extreme Condition User Facility (SECUF).

**Author contributions**

Y-C. L. and W. H. W. conceived the idea and supervised the project. L.X. S. carried out the preparation of the samples, and conducted the structure, magnetic and transport measurements. L.X.S. and X. L. fabricated the Hall devices. B. D., Y. Y., and H.L. performed the LTEM and STEM experiments. F. Z. and H. L. performed the ab initio calculation. All authors discussed the results and contributed to the manuscript preparation.

**Data Availability Statement**

The data that support the findings of this study are available from the corresponding author upon reasonable request.

# Reference

[1] L. Balents, Nature **2010**, 464, 199.

[2] Q. Chen, S. C. Bae, S. Granick, Nature **2011**, 469, 381.

[3] S. Yan, D. A. Huse, S. R. White, Science **2011**, 332, 1173.

[4] T. H. Han, J. S. Helton, S. Chu, D. G. Nocera, J. A. Rodriguez-Rivera, C. Broholm, Y. S. Lee, Nature **2012**, 492, 406.

[5] G. Xu, B. Lian, S. C. Zhang, Physical review letters **2015**, 115, 186802.

[6] H. Zhao, H. Li, B. R. Ortiz, S. M. L. Teicher, T. Park, M. Ye, Z. Wang, L. Balents, S. D. Wilson, I. Zeljkovic, Nature **2021**, 599, 216.

[7] L. Nie, K. Sun, W. Ma, D. Song, L. Zheng, Z. Liang, P. Wu, F. Yu, J. Li, M. Shan, D. Zhao, S. Li, B. Kang, Z. Wu, Y. Zhou, K. Liu, Z. Xiang, J. Ying, Z. Wang, T. Wu, X. Chen, Nature **2022**, 604, 59.

[8] K. Kuroda, T. Tomita, M. T. Suzuki, C. Bareille, A. A. Nugroho, P. Goswami, M. Ochi, M. Ikhlas, M. Nakayama, S. Akebi, R. Noguchi, R. Ishii, N. Inami, K. Ono, H. Kumigashira, A. Varykhalov, T. Muro, T. Koretsune, R. Arita, S. Shin, T. Kondo, S. Nakatsuji, Nature Materials **2017**, 16, 1090.

[9] Z. Hou, W. Ren, B. Ding, G. Xu, Y. Wang, B. Yang, Q. Zhang, Y. Zhang, E. Liu, F. Xu, W. Wang, G. Wu, X. Zhang, B. Shen, Z. Zhang, Advanced Materials **2017**, 29, 1.

[10] J. X. Yin, S. S. Zhang, H. Li, K. Jiang, G. Chang, B. Zhang, B. Lian, C. Xiang, I. Belopolski, H. Zheng, T. A. Cochran, S. Y. Xu, G. Bian, K. Liu, T. R. Chang, H. Lin, Z. Y. Lu, Z. Wang, S. Jia, W. Wang, M. Z. Hasan, Nature **2018**, 562, 91.

[11] E. Liu, Y. Sun, N. Kumar, L. Muchler, A. Sun, L. Jiao, S. Y. Yang, D. Liu, A. Liang, Q. Xu, J. Kroder, V. Suss, H. Borrmann, C. Shekhar, Z. Wang, C. Xi, W. Wang, W. Schnelle, S. Wirth, Y. Chen, S. T. B. Goennenwein, C. Felser, Nature Physics **2018**, 14, 1125.

[12] J.-X. Yin, S. S. Zhang, G. Chang, Q. Wang, S. S. Tsirkin, Z. Guguchia, B. Lian, H. Zhou, K. Jiang, I. Belopolski, N. Shumiya, D. Multer, M. Litskevich, T. A. Cochran, H. Lin, Z. Wang, T. Neupert, S. Jia, H. Lei, M. Z. Hasan, Nature Physics **2019**, 15, 443.

[13] P. P. Provenzano, Nat Mater **2020**, 19, 130.

[14] M. H. Christensen, X. Wang, Y. Schattner, E. Berg, R. M. Fernandes, Physics Review Letters **2020**, 125, 247001.

[15] X. Teng, L. Chen, F. Ye, E. Rosenberg, Z. Liu, J. X. Yin, Y. X. Jiang, J. S. Oh, M. Z. Hasan, K. J. Neubauer, B. Gao, Y. Xie, M. Hashimoto, D. Lu, C. Jozwiak, A. Bostwick, E. Rotenberg, R. J. Birgeneau, J. H. Chu, M. Yi, P. Dai, Nature **2022**, 609, 490.

[16] E. H. Hall, Proceedings of the Physical Society of London **1880**, 4, 325.

[17] C. M. Hurd, C. Chien, C. Westgate, Plenum, New York **1980**, 1.

[18] N. Nagaosa, J. Sinova, S. Onoda, A. H. MacDonald, N. P. Ong, Reviews of Modern Physics **2010**, 82, 1539.

[19] Y. Machida, S. Nakatsuji, S. Onoda, T. Tayama, T. Sakakibara, Nature **2010**, 463, 210.

[20] H. Ishizuka, Y. Motome, Physical Review B **2013**, 87, 081105(R).

[21] H. Chen, Q. Niu, A. H. MacDonald, Physics Review Letters **2014**, 112, 017205.

[22] J. Kübler, C. Felser, Europhysics Letters **2014**, 108, 67001.

[23] S. Nakatsuji, N. Kiyohara, T. Higo, Nature **2015**, 527, 212.

[24] M. Kimata, H. Chen, K. Kondou, S. Sugimoto, P. K. Muduli, M. Ikhlas, Y. Omori, T. Tomita, A. H. MacDonald, S. Nakatsuji, Y. Otani, Nature **2019**, 565, 627.

[25] T. Higo, K. Kondou, T. Nomoto, M. Shiga, S. Sakamoto, X. Chen, D. Nishio-Hamane, R. Arita, Y. Otani, S. Miwa, S. Nakatsuji, Nature **2022**, 607, 474.

[26] M. Ikhlas, S. Dasgupta, F. Theuss, T. Higo, S. Kittaka, B. J. Ramshaw, O. Tchernyshyov, C. W. Hicks, S. Nakatsuji, Nature Physics **2022**, 18, 1086.

[27] X. Li, J. Koo, Z. Zhu, K. Behnia, B. Yan, Nat Commun **2023**, 14, 1642.

[28] N. Kiyohara, T. Tomita, S. Nakatsuji, Physical Review Applied **2016**, 5, 064009.

[29] A. K. Nayak, J. E. Fischer, Y. Sun, B. Yan, J. Karel, A. C. Komarek, C. Shekhar, N. Kumar, W. Schnelle, J. Kübler, C. Felser, S. S. P. Parkin, Science Advances **2016**, 2, e1501870.

[30] M. Ikhlas, T. Tomita, T. Koretsune, M.-T. Suzuki, D. Nishio-Hamane, R. Arita, Y. Otani, S. Nakatsuji, Nature Physics **2017**, 13, 1085.

[31] T. Higo, H. Man, D. B. Gopman, L. Wu, T. Koretsune, O. M. J. van 't Erve, Y. P. Kabanov, D. Rees, Y. Li, M. T. Suzuki, S. Patankar, M. Ikhlas, C. L. Chien, R. Arita, R. D. Shull, J. Orenstein, S. Nakatsuji, Nature Photonics **2018**, 12, 73.

[32] C. Wuttke, F. Caglieris, S. Sykora, F. Scaravaggi, A. U. B. Wolter, K. Manna, V. Süss, C. Shekhar, C. Felser, B. Büchner, C. Hess, Physical Review B **2019**, 100, 085111.

[33] P. K. Rout, P. V. P. Madduri, S. K. Manna, A. K. Nayak, Physical Review B **2019**, 99.

[34] L. Xu, X. Li, L. Ding, K. Behnia, Z. Zhu, Applied Physics Letters **2020**, 117.

[35] M. Wu, H. Isshiki, T. Chen, T. Higo, S. Nakatsuji, Y. Otani, Applied Physics Letters **2020**, 116, 132408.

[36] T. Chen, T. Tomita, S. Minami, M. Fu, T. Koretsune, M. Kitatani, I. Muhammad, D. Nishio-Hamane, R. Ishii, F. Ishii, R. Arita, S. Nakatsuji, Nat Communications **2021**, 12, 572.

[37] Y. Pan, C. Le, B. He, S. J. Watzman, M. Yao, J. Gooth, J. P. Heremans, Y. Sun, C. Felser, Nature Materials **2022**, 21, 203.

[38] D. F. Zhang, J. K. Liang, Acta Physica Sinica **1966**, 22, 1005.

[39] E. Krén, G. Kádár, Solid State Communications **1970**, 8, 1653.

[40] H. Niida, T. Hori, Y. Nakagawa, Journal of the Physical Society of Japan **1983**, 52, 1512.

[41] H. Niida, T. Hori, Y. Yamaguchi, Y. Nakagawa, Journal of Applied Physics **1993**, 73, 5692.

[42] H. Niida, T. Hori, H. Onodera, Y. Yamaguchi, Y. Nakagawa, Journal of Applied Physics **1996**, 79.

[43] B. Balke, G. H. Fecher, J. Winterlik, C. Felser, Applied Physics Letters **2007**, 90.

[44] P. Kharel, Y. Huh, N. Al-Aqtash, V. R. Shah, R. F. Sabirianov, R. Skomski, D. J. Sellmyer, J Phys Condens Matter **2014**, 26, 126001.

[45] J. Z. Wei, R. Wu, Y. B. Yang, X. G. Chen, Y. H. Xia, Y. C. Yang, C. S. Wang, J. B. Yang, Journal of Applied Physics **2014**, 115.

[46] Z. Jiao, Z. Fu, J. Wang, R. Zhang, C. Jiang, Journal of Magnetism and Magnetic Materials **2019**, 489.

[47] L. X. Song, B. Ding, H. Li, S. H. Lv, Y. Yao, D. L. Zhao, J. He, W. H. Wang, Journal of Magnetism and Magnetic Materials **2021**, 536, 168109.

[48] L. X. Song, B. Ding, H. Li, S. H. Lv, Y. Yao, D. L. Zhao, J. He, W. H. Wang, Applied Physics Letters **2021**, 119, 152405.

[49] X. Zhao, J. Zhao, Advanced Materials Interfaces **2022**, 9, 2201606.

[50] L. Song, W. Li, S. Lv, X. Xi, D. Zhao, J. He, W. Wang, Journal of Applied Physics **2022**, 131, 173903.

[51] G. Kirste, J. Freudenberger, S. Wurmehl, Acta Materialia **2023**, 258, 119205.

[52] K. Minakuchi, R. Y. Umetsu, K. Ishida, R. Kainuma, Journal of Alloys and Compounds **2012**, 537, 332.

[53] Z. H. Liu, Y. J. Zhang, G. D. Liu, B. Ding, E. K. Liu, H. M. Jafri, Z. P. Hou, W. H. Wang, X. Q. Ma, G. H. Wu, Scientific Reports **2017**, 7, 515.

[54] H. Guo, Z. Feng, H. Yan, J. Liu, J. Zhang, X. Zhou, P. Qin, J. Cai, Z. Zeng, X. Zhang, X. Wang, H. Chen, H. Wu, C. Jiang, Z. Liu, Advanced Materials **2020**, 32, e2002300.

[55] Y. Zhang, Y. Sun, H. Yang, J. Železný, S. P. P. Parkin, C. Felser, B. Yan, Physical Review B **2017**, 95, 075128.

[56] V. Rai, S. Jana, M. Meven, R. Dutta, J. Perßon, S. Nandi, Physical Review B **2022**, 106, 195114.

[57] T. Nagamiya, S. Tomiyoshi, Y. Yamaguchi, Solid State Communications **1982**, 42, 385.

[58] S. Tomiyoshi, Journal of the Physical Society of Japan **1982**, 51, 803.

[59] L. Ye, M. Kang, J. Liu, F. von Cube, C. R. Wicker, T. Suzuki, C. Jozwiak, A. Bostwick, E. Rotenberg, D. C. Bell, L. Fu, R. Comin, J. G. Checkelsky, Nature **2018**, 555, 638.

[60] L. Ye, Y. Tian, X. Jin, D. Xiao, Physical Review B **2012**, 85 220403(R).

[61] T. Miyasato, N. Abe, T. Fujii, A. Asamitsu, S. Onoda, Y. Onose, N. Nagaosa, Y. Tokura, Physics Review Letters **2007**, 99, 086602.

[62] N. Manyala, Y. Sidis, J. F. DiTusa, G. Aeppli, D. P. Young, Z. Fisk, Nat Mater **2004**, 3, 255.

[63] S. Onoda, N. Sugimoto, N. Nagaosa, Physical Review B **2008**, 77, 165103.

[64] S. Onoda, N. Sugimoto, N. Nagaosa, Phys Rev Lett **2006**, 97, 126602.

[65] A. Sakai, Y. P. Mizuta, A. A. Nugroho, R. Sihombing, T. Koretsune, M.-T. Suzuki, N. Takemori, R. Ishii, D. Nishio-Hamane, R. Arita, P. Goswami, S. Nakatsuji, Nature Physics **2018**, 14, 1119.

[66] S. Nakatsuji, AAPPS Bulletin **2022**, 32.

[67] I. Belopolski, K. Manna, D. S. Sanchez, G. Chang, B. Ernst, J. Yin, S. S. Zhang, T. Cochran, N. Shumiya, H. Zheng, B. Singh, G. Bian, D. Multer, M. Litskevich, X. Zhou, S. M. Huang, B. Wang, T. R. Chang, S. Y. Xu, A. Bansil, C. Felser, H. Lin, M. Z. Hasan, Science **2019**, 365, 1278.

[68] D. F. Liu, A. J. Liang, E. K. Liu, Q. N. Xu, Y. W. Li, C. Chen, D. Pei, W. J. Shi, S. K. Mo, P. Dudin, T. Kim, C. Cacho, G. Li, Y. Sun, L. X. Yang, Z. K. Liu, S. S. P. Parkin, C. Felser, Y. L. Chen, Science **2019**, 365, 1282.

[69] G. Kresse, J. Furthmüller, Physical review B **1996**, 54, 11169.

[70] J. P. Perdew, K. Burke, M. Ernzerhof, Physical Review letters **1996**, 77, 3865.

[71] J. P. Perdew, K. Burke, M. Ernzerhof, Physical Review Letters **1998**, 80, 891.

[72] A. A. Mostofi, J. R. Yates, G. Pizzi, Y.-S. Lee, I. Souza, D. Vanderbilt, N. Marzari, Computer Physics Communications **2014**, 185, 2309.

[73] A. A. Mostofi, J. R. Yates, Y.-S. Lee, I. Souza, D. Vanderbilt, N. Marzari, Computer Physics Communications **2008**, 178, 685.

[74] Q. Wu, S. Zhang, H.-F. Song, M. Troyer, A. A. Soluyanov, Computer Physics Communications **2018**, 224, 405.